\documentclass[aps,prl,twocolumn,groupedaddress,letterpaper]{revtex4}
\usepackage[dvips]{graphics, color}
\usepackage{amssymb}
\usepackage{float}
\usepackage{bm,times}
\usepackage{graphicx}
\usepackage{textcomp}

\begin{document}
\title[Nanomechanical cooling with a microwave resonator]{Prospects for cooling nanomechanical motion by coupling to a superconducting microwave resonator}

\author{J. D. Teufel$^{1}$}
\author{C. A. Regal$^{1,2}$}
\author{K. W. Lehnert$^{1}$}
\email[E-mail: ]{john.teufel@colorado.edu}
\affiliation{$^{1}$
JILA, National Institute of Standards and Technology and the University of Colorado, and Department of Physics University
of Colorado, Boulder, Colorado 80309, USA}
\affiliation{$^{2}$Norman Bridge Laboratory of Physics 12-33, California Institute of Technology, Pasadena, California 91125}



\begin{abstract}
    Recent theoretical work has shown that radiation pressure effects can in principle cool a mechanical degree of freedom to its ground state.  In this paper, we apply this theory to our realization of an optomechanical system in which the motion of mechanical oscillator modulates the resonance frequency of a superconducting microwave circuit.  We present experimental data demonstrating the large mechanical quality factors possible with metallic, nanomechanical beams at 20~mK.  Further measurements also show damping and cooling effects on the mechanical oscillator due to the microwave radiation field.  These data motivate the prospects for employing this dynamical backaction technique to cool a mechanical mode entirely to its quantum ground state.
\end{abstract}

\maketitle

    Despite the recent progress in the fabrication, control, and measurement of macroscopic mechanical objects, a mechanical oscillator has yet to be observed in its motional ground state.  This long standing goal is a necessary prerequisite for many experiments that explore and make use of the quantum nature of motion, such as entanglement, squeezing, and quantum measurement.  Standard dilution refrigeration techniques can cool sufficiently high frequency mechanical resonators to their ground state; however these small, stiff oscillators are exceedingly difficult to measure with enough sensitivity to resolve the zero point motion. Furthermore, many current experimental realizations are incompatible with operation in a dilution refrigerator.  In these situations, feedback (either passive or active) seems to offer the most feasible route to ground state cooling.  In active feedback, a force is applied to the oscillator that is proportional to its velocity using the information acquired from a precise measurement of the mechanical motion.  This force can then damp the thermal, Brownian motion and cool the oscillator by coupling to the effectively ``colder" bath of the measurement.  In this way, the fundamental limit of this cooling technique is the sensitivity of the measurement.  In fact, it is only a Heisenberg-limited measurement that could completely cool to the quantum ground state \cite{Courty2001}.  So while this technique has been used to greatly reduce the effective temperature of a mechanical mode \cite{Arcizet2006,Poggio2007,Corbitt2007}, true ground state cooling is not a realistic possibility until a quantum limited measurement has been achieved.  

\begin{figure}\begin{center}\label{circuit}
\includegraphics[width=\linewidth]{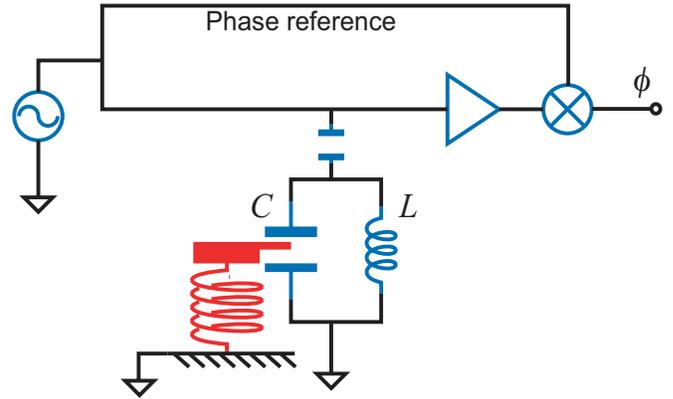}
\caption{Measurement schematic.  In this simplified circuit diagram, the microwave cavity is represented by an LC resonator capacitively coupled to a feedline.  The motion of a nanomechanical beam modulates this cavity resonance frequency by capacitively loading the circuit.  Thus, by measuring the phase, $\phi$, of the transmitted microwave signal via homodyne detection, the mechanical motion may be inferred.}
\end{center}\end{figure}

    While mesoscopic measurement techniques that can couple to lighter and smaller resonators have produced the closest approach to this measurement limit \cite{LaHaye2004,FJ2007}, technical issues have prevented measuring at the quantum limits of displacement sensitivity.  In contrast, the most precise measurements in term of absolute displacement sensitivity use interferometric techniques with visible light to infer the mechanical position \cite{Arcizet2006}.  This strong interaction between an electromagnetic resonance and mechanical motion offers another method for cooling.  This passive cooling can be achieved by engineering a system where the delayed radiation pressure of the light field serves to damp the oscillator's motion.  Recent experimental \cite{Schliesser2006,Naik2006,Brown2007,Thompson2007} and theoretical \cite{Marquardt2007,Wilson-rae2007} progress on the use of this dynamical back-action indicates that this method is poised to achieve true ground state cooling.  Specifically, Xue \textit{et al.} \cite{Xue2007} discuss cooling a mechanical degree of freedom by coupling to microwave cavity.  We have recently experimentally demonstrated such a system, whereby a nanomechanical beam is embedded in a superconducting, microwave cavity \cite{Regal2008}.  This paper further explores the prospects for backaction cooling for the experimental parameters feasible in our realization of an opto-mechanical system. 

\begin{figure*}\begin{center}\label{SEM}
\includegraphics[width=\linewidth]{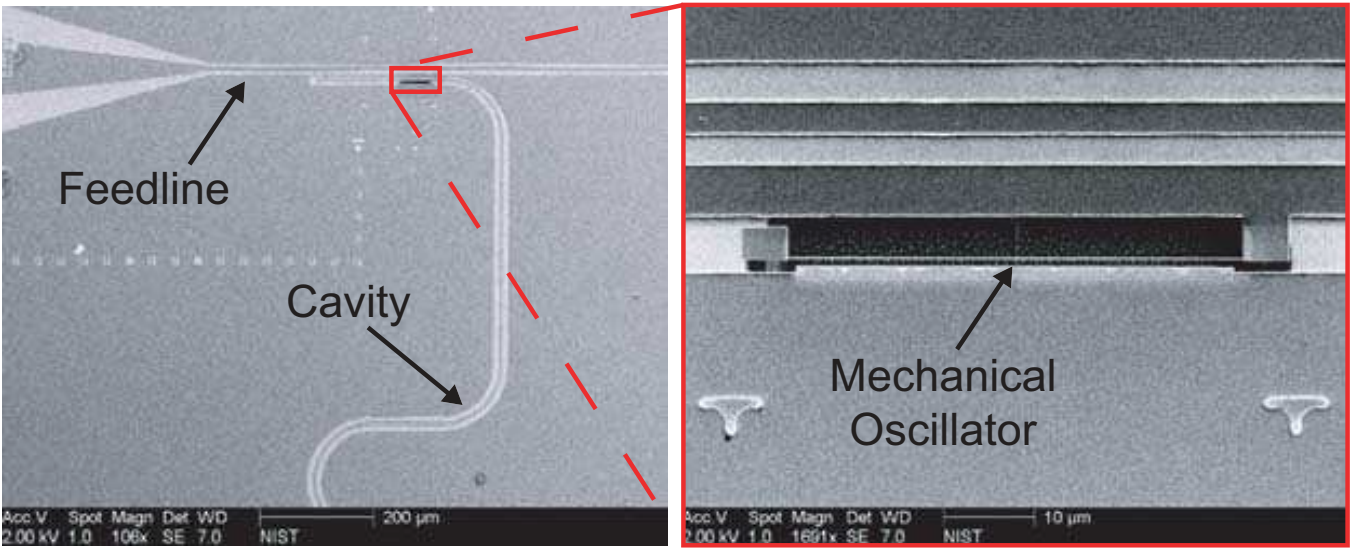}
\caption{Images of the device.  These scanning electron microscope images show both a portion of the microwave cavity (left) and the nanomechanical oscillator (right) fabricated out of aluminum.  The cavity consists of a $\lambda/4$ resonator that is capacitively coupled to a CPW feedline.  Within the voltage antinode of the cavity, a 50~$\mu$m long beam is suspended by etching away the silicon substrate.}
\end{center}\end{figure*}

    The Hamiltonian that describes an electromagnetic resonance whose resonant frequency is parametrically coupled to the motion of a mechanical oscillator is
\begin{eqnarray}
H &=& \hbar \omega_{0} \left(a^{\dagger}a+\case{1}{2}\right)+\hbar \omega_{m} \left(b^{\dagger}b+\case{1}{2}\right) \nonumber \\
&&-\hbar g a^{\dagger}a\left(b^{\dagger}+b\right)\delta x_{zp}.
\end{eqnarray}
Here $a^{\dagger}$ ($b^{\dagger}$) is the creation operator for the electromagnetic (mechanical) mode and $\omega_{0}$ ($\omega_{m}$) is the frequency of the unperturbed electromagnetic (mechanical) resonance. In the presence of a mechanical beam, the electromagnetic cavity resonance frequency is perturbed to $\omega_{c}$. The dispersive effect of a displacement $\hat{x} = (b^{\dagger} + b)\delta x_{zp}$ on the cavity resonance frequency is given by $g=-\frac{\partial\omega_{c}}{\partial x}$. This coupling implies that the light can be used to infer the displacement of a mechanically compliant object, and that photons in the cavity in turn exert a force on the mechanical oscillator.  If the mechanical resonance frequency is greater than or comparable to the energy decay rate of the cavity, $\gamma$, this force becomes substantially out of phase with the physical motion.  Recently, a fully quantum theory of this damping has shown that this effect is capable of cooling a mechanical mode to its ground state \cite{Marquardt2007,Wilson-rae2007,Genes2008,Grajcar2008}. 

    The mechanism for cooling via radiation damping can be understood as a Raman process whereby the parametric coupling allows drive photons to inelastically scatter to another frequency.  This process preferentially scatters photons to the cavity resonance frequency, as this is where the final photon density of states is maximal.  If an excitation frequency, $\omega_{e}$, is red detuned from cavity resonance by $\Delta=\omega_{e}-\omega_{c}<0$, the mechanical motion will be both damped and cooled.  In this paper, we will consider these quantum results within the weak-coupling limit \cite{Marquardt2007}, where the cavity linewidth is always greater than the mechanical line width.  This regime is sufficient for all experimental parameters considered here.  

    In general the radiation force is a complex quantity that can affect both the damping and the spring constant of the mechanical system. For an ideal, single-port cavity, the additional mechanical damping, $\Gamma$, and the mechanical frequency shift, $\delta\Omega$, due to the radiation field are given by
\begin{equation}\label{gamma} 
\Gamma=A\left[\frac{1}{\gamma^{2}+4\left(\Delta+\omega_{m}\right)^{2}}-\frac{1}{\gamma^{2}+4\left(\Delta-\omega_{m}\right)^{2}}\right]
\end{equation}
\begin{equation} 
\delta\Omega =\frac{A}{\gamma}\left[\frac{\Delta+\omega_{m}}{\gamma^{2}+4\left(\Delta+\omega_{m}\right)^{2}}+\frac{\Delta-\omega_{m}}{\gamma^{2}+4\left(\Delta-\omega_{m}\right)^{2}}\right]
\end{equation}
where 
\begin{equation}
 A=\frac{8g^{2}P\gamma^{2}}{m\omega_{m}\omega_{c}(\gamma^{2}+4\Delta^{2})}
\end{equation}
and $P$ is the incident power, $m$ is the effective mass, and $\gamma$ is the cavity linewidth \cite{Marquardt2007,Wilson-rae2007}.  

    In the absence of mechanical damping, the effective temperature of the mechanical degree of freedom is completely determined by the cavity-filtered photon shot noise.  Following the formalism of Marquardt $\textit{et al.}$ \cite{Marquardt2007}, the equilibrium number of thermal phonons, $\overline{n}_{p}$, for a given detuning, is completely determined by the ratio of $\omega_{m}$ and $\gamma$.
\begin{equation}\label{np} 
\overline{n}_{p}=-\frac{\gamma^{2}+4\left(\Delta+\omega_{m}\right)^{2}}{16 \omega_{m}\Delta}
\end{equation}    
    As $\omega_{m}$ becomes comparable to or greater than $\gamma$, one enters the so-called good cavity limit.  The quantity, $\overline{n}_{p}$, can be made less than unity only in this resolved sideband limit with a properly chosen red-detuned carrier.  The optimal detuning for the lowest effective temperature is given by
\begin{equation}
\Delta_{opt}=-\frac{1}{2}\sqrt{\gamma^{2}+4\omega_{m}^{2}}.   
\end{equation}
Thus, in the bad cavity limit ($\omega_{m}\ll\gamma$), $\Delta_{opt}\simeq-\gamma/2$ and $\overline{n}_{p}\simeq\gamma/(4 \omega_{m})\gg1$.  However, in the good cavity limit, $\Delta_{opt}\simeq-\omega_{m}$ and $\overline{n}_{p}\simeq\gamma^{2}/(16 \omega_{m}^{2})\ll1$.  It is important to note that, once in the good cavity limit, the frequency shift has zeros exactly at the points of maximal damping, $\Delta=\pm\omega_{m}$.  Thus, the effect of the radiation at these values of the detuning is purely a damping component.  

    In order to determine the final temperature in the presence of both a thermal reservoir and this effective photon temperature, one must weigh each temperature by the strength with which it is coupled to the mechanical oscillator.  Thus, the final steady-state occupancy of the mechanical oscillator, $\overline{n}_{f}$, is
\begin{equation}
\overline{n}_{f}=\frac{\Gamma \overline{n}_{p}+\gamma_{m}\overline{n}_{T}}{\Gamma+\gamma_{m}}.
\end{equation}
Here, $\gamma_{m}$ is the intrinsic mechanical damping, and $\overline{n}_{T}$ is the mean number of thermal phonons determined by the bath temperature.  As this final phonon number, $\overline{n}_{f}$, becomes less than unity, this mode of the mechanical oscillator asymptotically approaches its quantum mechanical ground state.  

    These equations may be summarized by noting that the requirements for reaching the ground state are threefold.  First, the intrinsic mechanical quality factor must be larger than the initial number of thermal phonons in the mechanical mode ($\overline{n}_{T}<Q_{m}$).  Second, the effective temperature of the photon bath must be sufficiently cold.  From equation \ref{np}, this implies that the system must be sufficiently in the good cavity limit ($\omega_{m}\gtrsim\gamma$).  The last requirement is that the optical damping is strong enough to overwhelm the coupling to the thermal bath ($\Gamma\gtrsim\gamma_{m}\overline{n}_{T}$).  Unlike optical systems, for which achieving the good cavity limit is the primary challenge, our microwave implementation naturally falls in this regime.  However, for our system, it is the last requirement on the coupling that sets the practical limit on temperatures attainable with passive cooling.  One additional caveat is that the ratio between the initial and final temperatures cannot exceed the ratio of the resonance frequency of the electromagnetic and mechanical modes ($\overline{n}_{T}/\overline{n}_{f}\leq\omega_{c}/\omega_{m}$) \cite{Grajcar2008}.  While this restriction could in principle preclude ground state cooling, it is not a limiting factor for the experimental parameters considered in this paper.  

\begin{figure}\begin{center}\label{Qvsf}
\includegraphics[width=\linewidth]{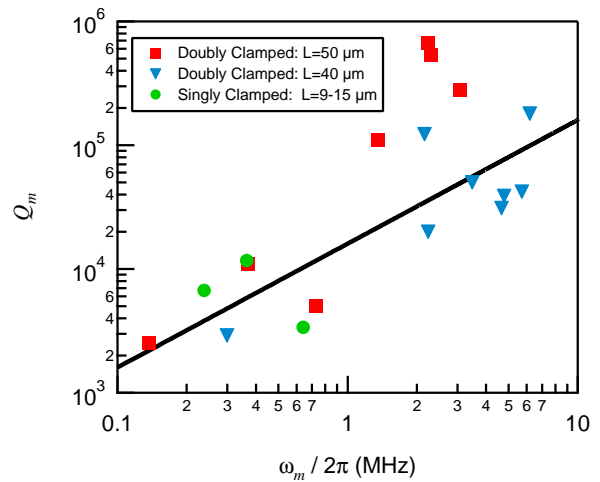}
\caption{Mechanical resonance frequencies and quality factors.  This plot shows 18 mechanical oscillators that have been measured with our microwave interferometric technique at a cryostat temperature of 20~mK. For the doubly clamped geometry, film tension can be used to greatly increase both resonance frequency and the the quality factor of the mechanical mode.  The black line shows the $Q_{m}$ one would expect if the beam tension were only changing the spring constant and leaving the mechanical damping, $\gamma_{m}$, constant.}
\end{center}\end{figure}

    In our system, the motion of a freely-suspended, nanomechanical beam modulates the resonance frequency of a superconducting microwave cavity.  As shown in figure 2, this cavity takes the form of a $\lambda/4$ coplanar waveguide guide (CPW) resonator that is capacitively coupled to a feedline.  Such CPW resonators simultaneously provide many desirable characteristics, and have been utilized in several recent experiments in fields such as radio-astronomy \cite{Day2003,Mates2008} and quantum computing \cite{Wallraff2004}.  These include ease of fabrication, cryogenic operation, small mode volume and low loss ($Q>10^{5}$).  Because the mode volume is so small, even the nanomechanical motion of picogram oscillators can significantly influence the electromagnetic resonance.     
    
    A high aspect-ratio, nanomechanical beam is placed at a voltage antinode within the cavity.  In this configuration, the motion of the beam gives rise to a phase shift in the transmitted microwave signal as shown in Figure 1.  Both the beam and the CPW structures are fabricated with a single layer of evaporated aluminum.  Typical cavity frequencies range from 4 to 10~GHz.  As dynamical backaction cooling relies on the cavity being excited with a purely coherent excitation, the thermal population of the cavity modes must be much less than unity.  For microwave cavities, this condition is realized by operating in a dilution refrigerator.  The coupled $Q$ of the cavity is designed to be roughly $10^{4}$ so that internal losses do not dominate.  While these $Q$ values are modest compared to optical Fabry-Perot cavities used to measure micromechanical oscillators, it is the cavity linewidth that determines if the system is in the good cavity limit.  Thus, by using microwave frequencies and superconducting resonators, linewidths can be made less than 1~MHz.  

    Initial measurements have yielded very encouraging results for the low temperature mechanical properties of beams made entirely out of evaporated aluminum \cite{Regal2008}.  Figure 3 shows the range of mechanical resonance frequencies and quality factors for devices at a cryostat temperature of 20~mK.  These measurements include both doubly-clamped beams and cantilever style oscillators.  These beams range in length from 9 and 50~$\mu$m and have a height and width of roughly 100~nm.  Some of the doubly-clamped beams were thermally annealed in order to place them under tensile stress.  Tension not only increases the mechanical resonance frequency, but also greatly increases the mechanical quality factor as well.  This effect is analogous to Fabry-Perot experiments in which the optical spring effect is exploited to increase the mechanical stiffness without changing the damping \cite{Corbitt2007}.  Similar results have recently been observed in silicon nitride beams and membranes where quality factors exceed $10^{6}$ at room temperature \cite{Verbridge2008} and $10^{7}$ at 300~mK \cite{Zwickl2008}.  While our beams without tension tend to have quality factors less than $10^{4}$, tension reproducibly results in oscillators with $Q>10^{5}$.    

\begin{figure}\begin{center}\label{QvsDetune}
\includegraphics[width=\linewidth]{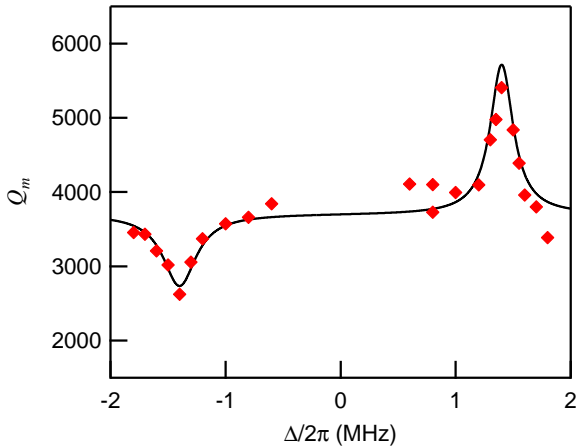}
\caption{Detuning dependence of the mechanical quality factor.  The mechanical quality factor, $Q_{m}$, displays the expected effects of radiation damping.  As the microwave carrier is red (blue) detuned from the cavity resonance by $\pm\omega_{m}$, the dynamical backaction results in a maximal decrease (increase) in $Q_{m}$. The width of these peaks in $Q_{m}$ is set by the cavity linewidth, $\gamma$=2$\pi\times$300~kHz.  For this device, g=2$\pi\times$11~kHz/nm, $\omega_{m}$=2$\pi\times$1.41~MHz.  These data are taken at a cryostat temperature of 100~mK and a constant circulating microwave power of 1~$\mu$W.}
\end{center}\end{figure}

    From equation \ref{np}, one can show that passively cooling to a thermal occupancy of less than one phonon is theoretically possible for a ratio of $\omega_{m}/\gamma>1/\sqrt{32}$.  This requirement is easily met for all beams and cavity parameters shown above.  The limit of the cooling in our system is not the effective temperature of the radiation effects ($\textit{i.e.}$ reaching the good cavity limit), but rather how strongly the radiation can be coupled to the mechanical motion.  It is only if the radiation damping, $\Gamma$, is much larger than the intrinsic damping, $\gamma_{m}$, that the mechanical oscillator decouples from its thermal environment.  In theory, weak coupling can be overcome by increasing the circulating photon power.  In practice, this quantity is limited by the critical current of the superconducting film.  For our aluminum cavities, the microwave resonance becomes non-linear at a circulating power of order 1~$\mu$W.  

\begin{figure}\begin{center}\label{cooling}
\includegraphics[width=\linewidth]{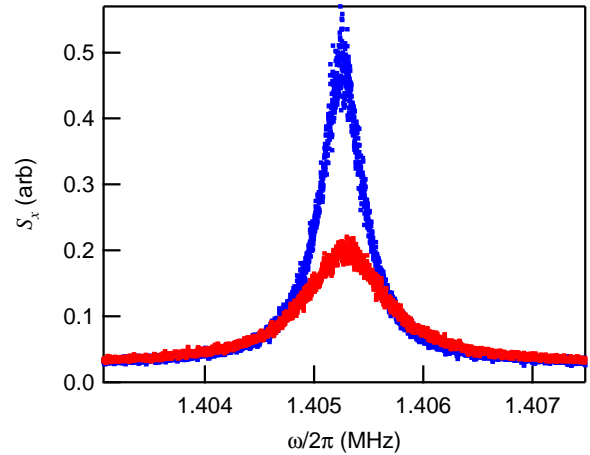}
\caption{Thermal noise of a mechanical resonance with red and blue detuning.  For these curves, the microwave carrier is detuned by the mechanical resonance frequency so that the dynamical backaction has a maximal effect.  The blue curve is when the microwave carrier is blue detuned from the cavity resonance, increasing both $Q_{m}$ and the rms motion of the oscillator.  Likewise, the red curve is red detuned, and the nanomechanical beam experiences damping and cooling.  These data are taken with the cryostat held at a temperature of 200~mK.  The 50$\%$ reduction in the area of the thermal peaks with blue and red detuning represents and effective temperature of 280 and 140~mK, respectively.}
\end{center}\end{figure}

    One device showed coupling that is an order of magnitude greater than expected.  Later SEM imaging indicated that this is due to part of a doubly clamped beam actually touching the adjacent ground plane.  While this configuration makes it difficult to know the precise mode shape or effective mass, the driven response shows a clear mechanical resonance at 1.41~MHz with a quality factor of 3800.  Due to this large coupling, there is clear evidence of dynamical backaction in both the quality factor and effective temperature of the beam even at low microwave powers.  For this device the cavity resonance frequency is 7.55~GHz with the quality factor of 25,000.  Figure 4 shows the mechanical quality factor as a function of microwave carrier detuning while maintaining a constant circulating power.  This experiment is well in the good cavity limit ($\omega_{m}=4.7\times\gamma$) as is immediately evident from figure 4.  There is a sharp peak (dip) in the mechanical quality factor when the carrier is blue (red) detuned by the mechanical resonance frequency.  The black line shows a fit to equation \ref{gamma}, which reproduces the independently determined values for $\omega_{m}$ and $\gamma$.     
    
    This change in the mechanical quality factor implies that the thermal motion of the beam should also be affected.  In order to measure this effect, the cryostat is held a temperature of 200~mK so that the thermal motion of the beam is clearly visible.  As shown in figure 5, the spectral density is measured as a function of frequency with the carrier both red and blue detuned by the mechanical resonance frequency.  The area of the Lorentzian peak is proportional to the rms motion of this mode of the beam, and hence its effective temperature.  These data show a 50$\%$ reduction in this area between the blue and red detuning.  Thus, the for the blue detuning, the beam is coupled to a negative temperature bath (as given by equation \ref{np}) and the temperature increases from 200~mk to 280~mK.  Analogously, in the case of red detuning, the beam cools to an effective temperature of 140~mK.  While this level of cooling is far from what is required for ground state cooling, it clearly demonstrates the dynamical backaction of the microwave cavity on the temperature of the mechanical oscillator.  As recent data has shown that mechanical quality factors of greater than $10^{5}$ are feasible, it is interesting to note that this would imply a hundred fold increase in the cooling effect for similar radiation damping.  It is with a device that possesses both a high mechanical quality factor and the large coupling seen in this particular measurement that ground state cooling becomes a foreseeable prospect.   

\begin{figure}\begin{center}\label{nf}
\includegraphics[width=\linewidth]{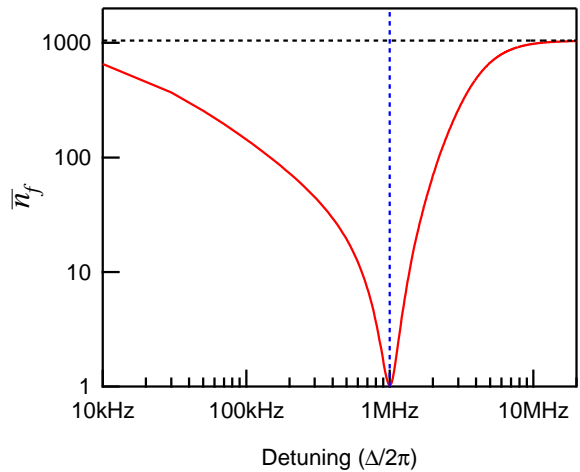}
\caption{Mechanical oscillator phonon number as a function of red detuning for possible future experimental parameters.  By coupling a 1~MHz mechanical resonator with a $Q_{m}$ of 5$\times 10^{5}$ to a microwave cavity resonance with a frequency of 10~GHz and $Q$ of 40,000, dramatic cooling is possible.  This assumes coupling of $g=2\pi\times$80~kHz/nm and a circulating power of 2~$\mu$W that are plausible with future improvements. At the optimal detuning ($\Delta=-\omega_{m}$), the oscillator cools from more than 1000 phonons to approximately one.}
\end{center}\end{figure}   

    In future experiments, we believe such large coupling strengths should be reproducibly achievable by implementing several improvements.  The relative coupling constant, $g/\omega_{c}$, depends on the $\textit{fractional}$ change in capacitance for a given mechanical displacement.  Thus, the beam's effect on the cavity (and hence the cavity's effect of the beam) may be maximized in two ways.  The first method is to increase the mutual capacitance between the mechanical oscillator and the cavity.  While this can be accomplished by simply fabricating the suspended beam closer to the adjacent ground plane, this separation is limited, in a practical sense, by intrinsic floppiness of the mechanical oscillator.  A more feasible option is to increase the length of the beam while using tension to obtain the desired mechanical resonance frequency.  A second method for improving the coupling involves reducing the total capacitance of the electromagnetic resonance.  This may be accomplished by increasing either the resonance frequency ($\omega_{c}\simeq(LC)^{-1/2}$) or the characteristic impedance ($Z_{0}\simeq\sqrt{L/C}$) of the cavity.  With all of these improvements, it is foreseeable that this total capacitance can be reduced by more than an order of magnitude from our current design \cite{Regal2008}.  

    To quantify the cooling potential of dynamical backaction, consider an optimized device with the parameters listed in the caption of figure 6.  This shows the theoretical final phonon occupancy as a function of red detuning.  Here, a 1~MHz mechanical mode begins in thermal equilibrium with a dilution refrigerator at 50~mK ($\sim$1000 phonons).  As the detuning reaches the optimal value, the beam cools by the dynamical backaction to less than 50~$\mu$K (1 phonon).  This assumes that the coupling strength is greatly increased by making a longer nanomechanical beam (200~$\mu$m) and by increasing the characteristic impedance of the microwave resonator to 300~$\Omega$.  While similar results could be achieved with a range of experimental parameters, this example serves to demonstrate the prospects for passive cooling with realistic parameters.  

    In conclusion, we have applied the theoretical analysis of dynamical backaction to our microwave cavities to show that this technique is a feasible option for cooling a mechanical mode to its quantum ground state.  These theoretical expectations are framed within the context of currently achievable experimental parameters.  Our measurements have demonstrated nanomechanical oscillators with resonance frequencies between 0.1 and 6~MHz, exhibiting high quality factors at cryogenic temperatures.  Furthermore, preliminary experimental results on a system in the good cavity limit demonstrate radiation damping and cooling effects.  Together, these results indicate that the appropriate combination of experimentally demonstrated parameters will allow cooling below dilution refrigerator temperatures, perhaps even to the ground state of the mechanical oscillator. 

    The authors thank D. R. Schmidt for taking the scanning electron micrographs in figure 2.
\bibliographystyle{apsrev}
\bibliography{Teufel_bib}
\end{document}